\documentclass[12pt,a4paper]{iopart}

\usepackage{graphicx}

\begin{document}

\title{Heavy flavour jet abundances and tagging rates analytically determined}

\author{L. Sonnenschein}
\address{RWTH Aachen, III. Phys. Inst. A}
%\ead{\mailto{Lars.Sonnenschein@cern.ch}}

\begin{abstract}
Heavy flavour jet tagging is widely used in the determination of
cross sections including the production of heavy flavoured quarks.
This requires the knowledge of heavy and light
flavour jet tagging efficiencies and their uncertainties.   
A system of eight non-linear equations can be used to determine
these quantities by means of two different tagging algorithms 
and two data sets which differ in heavy flavour jet content.
The analytical solution of this system of equations, derived by means of resultants
is described in detail, including a discussion about its singularities
which provide important insights, such as prescriptions to prevent badly chosen sample 
flavour compositions and working points of the used tagging algorithms.
The analytical solution also provides an efficient and transparent way to determine 
the uncertainties on the solved quantities, taking correlations into account. 
\end{abstract}

%{\it Key words:} Tagging, Heavy flavour jets \newline
\noindent{\it Heavy flavour jets, Tagging \/}

%{\it PACS: 14.65.Fy, 14.65.Bt, 21.10.Tg}

%\submitto{\JPA}
%\maketitle

\section{Introduction}

Heavy flavour jet tagging is widely used in the determination of
cross sections including the production of heavy flavoured quarks.
This encompasses interesting physics processes, such as 
top anti-top quark pair production, $Z\rightarrow b\bar{b}$ production,
Higgs boson production ($H\rightarrow b\bar{b}$), etc.
The $b$ quarks fragment into long-lived hadrons,
whose decay products manifest typically in the tracking system of a 
%collider 
detector as displaced charged particle tracks. 
The energy deposits of particles are measured in the calorimeter of the detector. 
They are used to reconstruct jets, to which charged tracks can be associated. 

In simulation one can easily identify the quark flavour
of a jet by checking if a heavy flavour quark or hadron is geometrically 
located inside the jet. In data one has to rely on reconstructed objects and quantities
such as associated track displacements to determine the flavour content of a jet.
Dedicated algorithms - referred to as tagging algorithms - 
have been developed to determine the flavour of jets based on their lifetime information.
This procedure is not perfect and implies certain probabilities
for the correct identification of jet flavours.   
The uncertainties on these probabilities enter among others into the systematics of 
cross section measurements of the physics processes mentioned above.
Therefore it is indispensable to determine the heavy and light
flavour jet tagging efficiencies and their uncertainties as accurate as possible. 
  
System8 
%\cite{Clement2003} 
%\cite{Clement2004} 
\cite{Clement2006} 
provides an Ansatz which allows one
to determine these quantities.
It consists of a system of eight algebraic equations
\begin{eqnarray} \label{system8} \nonumber
  n_{\phantom{AB}} & = & \; n_b + n_q , \\ \nonumber
  n_{T\phantom{B}} & = & \; \epsilon_T n_b + f_T n_q , \\ \nonumber 
  n_{S\phantom{B}} & = & \; \epsilon_S n_b + f_S n_q , \\ \nonumber 
  n_{TS} & = & \epsilon_T \; \epsilon_S n_b + f_T f_S n_q , \\
  p_{\phantom{AB}} & = & \; p_b + p_q , \\ \nonumber
  p_{T\phantom{B}} & = & \; \epsilon_T p_b + f_T p_q , \\ \nonumber 
  p_{S\phantom{B}} & = & \; \epsilon_S p_b + f_S p_q , \\ \nonumber 
  p_{TS} & = & \; \epsilon_T \epsilon_S p_b + f_T f_S p_q  , \\ \nonumber 
\end{eqnarray}
where $n$ is a data sample of a given number of jets.
It consists of two subsamples, one containing the number of 
heavy flavour jets $n_b$ and another one containing the number of light flavour jets $n_q$.
The contribution of charm flavoured jets 
is predominantly absorbed into the light flavour subsample.
Similarly, another data sample $p$ can be subdivided in its
heavy and light flavour contributions $p_b$ and $p_q$. It is important that the two
data samples $n$ and $p$
differ in their flavour composition; 
otherwise the system of eight equations degenerates to a system of
four linearly independent equations which cannot be solved for its eight unknowns
anymore.
The two data samples can share common jets 
and are therefore in general not statistically independent.
The efficiencies to tag heavy flavour jets are designated as $\epsilon$ and the
efficiencies to tag light flavour jets - also referred to as fake rates -  
are designated with $f$. 
To determine these efficiencies and fake rates, two different tagging algorithms
$T$ and $S$, which have to be de-correlated \cite{Scanlon2006}, are needed. 

The de-correlation requirement can be achieved in choosing two tagging algorithms $T$ 
and $S$, which determine the flavour based on the information of different
jet or associated track quantities which are de-correlated.
In practice one tagging algorithm can be chosen,
based on the lifetime information of a jet through the impact parameter of associated tracks, e.g. by means of a secondary vertex \cite{Abazov2006} \cite{Sonnenschein2003}
%or a jet lifetime probability \cite{Clement2002} \cite{Clement2003b} 
tagger. 
Since muons in jets are a signature of heavy flavour decay
an appropriate choice for the second tagging algorithm is a muon tagger, 
which looks for muons in jets with a certain transverse 
momentum with respect to the jet axis. 
%It has to be verified by means of simulation that the 
%correlation factors are consistent with one. 
In the following, it is assumed that the taggers are fully de-correlated, 
in which case the unknown quantities can be obtained 
relying on data only. Otherwise, correlation factors determined by simulation 
have to be introduced into the system of eqs. \ref{system8}. 
%In this case some terms do not cancel out and the resultant yields
%a univariate polynomial of degree eight, which can not be solved analytically anymore
%(though it could be solved semi-analytical by means of Sturm's theorem which gives the number
%of real roots of a given univariate polynomial of arbitrary degree in a given interval.
%Finally one would need to polish the roots e.g. by binary bracketing and take care about the
%solution ambiguities)

The known quantities that appear in the system of equations (the number of jets in each of the two
samples before and after one or both taggers have been applied) are on the left hand side of eqs.
\ref{system8}. The eight unknown quantities, which are the heavy and light flavour jet content 
of the samples and the tagging efficiencies and fake rates, appear on the right hand side.

The analytical solution of the system of equations will be derived in the next section, 
followed by a discussion of its numerical stability and the determination of errors
on the solved quantities in section \ref{errors}.
An explicit example, showing the performance of the method is given in section
\ref{performance}.

\section{Analytical solution}

Elementary algebraic operations permit one to reduce the system of eight equations
to two equations of two unknowns, which are polynomials of multi-degree two. 
The choice of the two unknowns is not unique.  
Since the primary goal is to determine the heavy flavour jet
tagging efficiency and the fake rate of a tagging algorithm $T$ to be used in various 
data analyses, it is convenient to keep the two unknowns $\epsilon_T$ and $f_T$ in the 
two polynomials. The other unknowns can be obtained by backward substitution.
In general two multi-variate polynomials of two unknowns with arbitrary degree can be solved
by means of resultants \cite{Sonnenschein2005} \cite{Sonnenschein2006}.
The two polynomials can be written in the form
\begin{eqnarray} \label{poly2} \nonumber
  r_1 & := & a_0f_T+a_1 \; = \; 0 , \\  
  r_2 & := & b_0f_T+b_1 \; = \; 0 . 
\end{eqnarray}
The coefficients $a_0$, $a_1$ and $b_0$, $b_1$ are polynomials in the unknown $\epsilon_T$.
Their explicit expressions are given by
\begin{eqnarray} \nonumber
 a_0 & = & a_{00} \epsilon_T + a_{01} , \\ \nonumber 
 a_1 & = & a_{10} \epsilon_T + a_{11} , \\ 
 b_0 & = & b_{00} \epsilon_T + b_{01} , \\ \nonumber
 b_1 & = & b_{10} \epsilon_T + b_{11} ,
\end{eqnarray}
where the coefficients can be expressed in terms of the initial known quantities as
\begin{eqnarray} \nonumber
 a_{00} & = & -n^2 p_S + n_S n p , \\ \nonumber
 a_{01} & = & n n_T p_S - n_S n_T p , \\ 
 a_{10} & = & a_{01} , \\ \nonumber
 a_{11} & = & n_T n_{TS} p - n_{TS} p_T n + n_S n_T p_T -n_T^2 p_S
\end{eqnarray}
for the first polynomial and
\begin{eqnarray} \nonumber
 b_{00} & = & -n_S n_T p - p_{TS} n^2 + n_{TS} p n + n_S p_T n , \\ \nonumber
 b_{01} & = & -n_{TS} p_T n + p_{TS} n n_T , \\ 
 b_{10} & = & b_{01} , \\ \nonumber
 b_{11} & = & n_{TS} p_T n_T - p_{TS} n_T^2  
\end{eqnarray}
for the second polynomial.
The resultant with respect to $f_T$ in eqn. \ref{poly2} can then be obtained by equating the determinant 
of the Sylvester matrix \cite{Akritas1989}
\begin{equation}
  Res(f_T)=\left| 
   \begin{array}{cc}
     a_0 & b_0 \\
     a_1 & b_1
   \end{array}
 \right| 
\end{equation}
to zero. The solution is a univariate polynomial of degree two which can be 
solved for the first unknown 
\begin{equation} \label{epsilonTsol}
  \hspace*{-5ex}
  \epsilon_T  = -\frac{1}{2}\frac{a_{00}b_{11}-a_{11}b_{00}}{a_{00}b_{10}-a_{10}b_{00}} + \sqrt{ \left(-\frac{1}{2}\frac{a_{00}b_{11}-a_{11}b_{00}}{a_{00}b_{10}-a_{10}b_{00}}\right)^2-\frac{a_{01}b_{11}-a_{11}b_{01}}{a_{00}b_{10}-a_{10}b_{00}}  } .
\end{equation}
The solution has a two-fold ambiguity originating from 
the initial system of equations which is symmetric with respect to exchange of efficiencies
and fake rates. In practice, the larger solution given above (positive square root) 
corresponds to the efficiency. 
The remaining unknowns can be obtained through backward substitution
and are given by
\begin{equation}
  f_T =  -\frac{a_1}{a_0} ,
\end{equation}

\begin{equation} \label{epsilonSsol}
 \epsilon_S = \frac{n_{TS}-n_S f_T}{n_T - f_T n} ,
\end{equation}

\begin{equation} \label{fakeS}
 f_S = \frac{n_S (\epsilon_T-f_T) + \epsilon_S (f_t n - n_T)}{n  \epsilon_T - n_T} ,
\end{equation}

\begin{equation} \label{sample_nb}
  n_b =  \frac{n_T - f_T n}{\epsilon_T - f_T} ,
\end{equation}

\begin{equation}
  n_q =  n - n_b ,
\end{equation}

\begin{equation} \label{sample_pb}
  p_b =  \frac{p_T - f_T p}{\epsilon_T - f_T} 
\end{equation}
and
\begin{equation}
  p_q =  p - p_b .
\end{equation}
In the more general case where arbitrary correlation factors are taken into account in the system of 
equations some terms do not cancel out each other and the resultant yields
a univariate polynomial of degree eight, which can not be solved analytically anymore
(though it could be solved semi-analytical by means of Sturm's theorem which gives the number
of real roots of a given univariate polynomial of arbitrary degree in a given interval \cite{Sonnenschein2005}.
The roots would need to be polished e.g. by binary bracketing and one would need to take care about the
increased number of solution ambiguities).

\section{Numerical stability and error propagation}
\label{errors}

The solution of the initial system of eight equations contains three singularities due to
vanishing denominators in case of not carefully chosen working points.
A first singularity occurs in the denominator of eqn. \ref{epsilonSsol} which determines the efficiency $\epsilon_S$. The singularity of the type ($n_T-n f_T$) appears if
all tagged jets are of light flavour and they have been misidentified as heavy flavour jets.
This would certainly correspond to an ill-posed 
initial condition which prohibits the sensible use of
a given tagging algorithm.
The second singularity occurs in the denominator of eqn. \ref{fakeS} which determines the fake 
rate $f_S$. A singularity of type ($n_T-n\epsilon_T$) appears if
all tagged jets are of heavy flavour. 
This particular case corresponds to a perfect algorithm 
which could be solved with just four equations. In practice it cannot be avoided that some 
of the tagged jets will be misidentified. The fake rate of optimised tagging algorithms
depends on the working point, which is typically chosen at the percent level ( see e.g. 
\cite{Abazov2005}) 
far away from zero within real precision at which point the implementation of the 
analytical solution looses its power of prediction.
The last singularity occurs in the denominator of eqs. \ref{sample_nb} and \ref{sample_pb}. 
There, the singularity of type ($\epsilon_T-f_T$) corresponds
to the case of equal efficiency and fake rate. The efficiency and fake rate of a given tagging
algorithm could be determined with a system of four equations but one would not be able 
to determine the heavy and light flavour content of the sample. This working point
is far away from typical chosen settings which are driven by the goal of having a maximal
efficiency at minimal fake rate. 
To prevent the problematic regions around the singularities one has to choose working points
of the taggers which avoid small or even vanishing denominators.
This can be achieved in probing different working points and verifying the values 
of the denominators.

The errors of the analytical solution are obtained through gaussian error propagation
taking all correlations into account.
The evaluation of the errors of subsequently solved quantities, 
obtained through backward substitution,
is accurately done by error propagation through differentials as explained below. 
In this way
the errors of the subsequent solved quantities conserve all their dependencies
on the knowns such that terms with different signs compensate before the gaussian 
quadrature takes place.
%This has been done in contrast to the possibility of keeping the final solved 
%quantity (here $\epsilon_T$) and its error in the Gaussian error propagation of a 
%subsequent quantity, preventing possible cancellations since added in quadrature 
%without former substitution. 

The errors of the knowns entering into the error propagation are the errors 
of the data samples $n$ and $p$, which are assumed to be Poisson distributed.
The errors of the tagged subsamples ($n_T$, $p_T$, $n_S$, $p_S$, $n_{TS}$ and $p_{TS}$)
are taken binomial since the subsamples have been obtained by applying a binary 
acceptance/rejection procedure through the taggers.
They can be computed from the known variables as
\begin{eqnarray} \label{knownerrors} \nonumber
  \Delta n_{\phantom{AB}}& = & \sqrt{n} , \\ \nonumber
  \Delta n_{T\phantom{B}} & = & \sqrt{n_T  (n-n_T)/n} , \\ \nonumber
  \Delta n_{S\phantom{A}} & = & \sqrt{n_S (n-n_S)/n} , \\
  \Delta n_{TS} & = & \sqrt{n_{TS} (n-n_{TS})/n} \\ \nonumber
%  \\ \nonumber
%  \Delta p_{\phantom{AB}} & = & \sqrt{p} \\ \nonumber
%  \Delta p_{T\phantom{B}} & = & \sqrt{p_T (p-p_T)/p} \\ \nonumber
%  \Delta p_{S\phantom{a}} & = & \sqrt{p_S (p-p_S)/p} \\ \nonumber
%  \Delta p_{TS} & = & \sqrt{p_{TS} (p-p_{TS})/p} \\ \nonumber
\end{eqnarray}
and equivalently for the data sample $p$.
The variance of the tagging efficiency is then given by
\begin{eqnarray} \label{et_err2} \nonumber
  \hspace*{-8ex}
  ( \Delta \epsilon_T )^2 & = & \left(\frac{d\epsilon_T}{d n}\right)^2 (\Delta n)^2 
  + \left(\frac{d\epsilon_T}{d p}\right)^2  (\Delta p)^2 
  + \left(\frac{d\epsilon_T}{d n_T}\right)^2  (\Delta n_T)^2 \\ \nonumber
  & & + \left(\frac{d\epsilon_T}{d p_T}\right)^2 (\Delta p_T)^2 
      + \left(\frac{d\epsilon_T}{d n_S}\right)^2 (\Delta n_S)^2 
      + \left(\frac{d\epsilon_T}{d p_S}\right)^2 (\Delta p_S)^2 \\ \nonumber
    & & + \left(\frac{d\epsilon_T}{d n_{TS}}\right)^2 (\Delta n_{TS})^2
    + \left(\frac{d\epsilon_T}{d p_{TS}}\right)^2 (\Delta p_{TS})^2
    + 2 \frac{d\epsilon_T}{d n} \frac{d\epsilon_T}{d p} \varrho_{np} \Delta n \Delta p \\ \nonumber
    & & + 2 \frac{d\epsilon_T}{d n} \frac{d\epsilon_T}{d p_T} \varrho_{np_T} \Delta n \Delta p_T
    + 2 \frac{d\epsilon_T}{d n} \frac{d\epsilon_T}{d p_S} \varrho_{np_S} \Delta n \Delta p_S \\ \nonumber 
\\[-1.5ex] \nonumber
    & & + 2 \frac{d\epsilon_T}{d n} \frac{d\epsilon_T}{d p_{TS}} \varrho_{np_{TS}} \Delta n \Delta p_{TS} 
    + 2 \frac{d\epsilon_T}{d n_T} \frac{d\epsilon_T}{d p} \varrho_{n_Tp} \Delta n_T \Delta p \\ \nonumber
    & & + 2 \frac{d\epsilon_T}{d n_T} \frac{d\epsilon_T}{d p_T} \varrho_{n_Tp_T} \Delta n_T \Delta p_T
    + 2 \frac{d\epsilon_T}{d n_T} \frac{d\epsilon_T}{d p_S} \varrho_{n_Tp_S} \Delta n_T \Delta p_S \\ \nonumber
    & & + 2 \frac{d\epsilon_T}{d n_T} \frac{d\epsilon_T}{d p_{TS}} \varrho_{n_Tp_{TS}} \Delta n_T \Delta p_{TS}
        + 2 \frac{d\epsilon_T}{d n_S} \frac{d\epsilon_T}{d p} \varrho_{n_Sp} \Delta n_S \Delta p \\ \nonumber
    & & + 2 \frac{d\epsilon_T}{d n_S} \frac{d\epsilon_T}{d p_T} \varrho_{n_Sp_T} \Delta n_S \Delta p_T %\\ \nonumber
    + 2 \frac{d\epsilon_T}{d n_S} \frac{d\epsilon_T}{d p_S} \varrho_{n_Sp_S} \Delta n_S \Delta p_S \\ \nonumber
    & & + 2 \frac{d\epsilon_T}{d n_S} \frac{d\epsilon_T}{d p_{TS}} \varrho_{n_Sp_{TS}} \Delta n_S \Delta p_{TS}
    + 2 \frac{d\epsilon_{T}}{d n_{TS}} \frac{d\epsilon_T}{d p} \varrho_{n_{TS}p} \Delta n_{TS} \Delta p \\ \nonumber
    & & + 2 \frac{d\epsilon_{T}}{d n_{TS}} \frac{d\epsilon_T}{d p_T} \varrho_{n_{TS}p_T} \Delta n_{TS} \Delta p_T 
    + 2 \frac{d\epsilon_{T}}{d n_{TS}} \frac{d\epsilon_T}{d p_S} \varrho_{n_{TS}p_S} \Delta n_{TS} \Delta p_S \\ \nonumber
    & & + 2  \frac{d\epsilon_{T}}{d n_{TS}} \frac{d\epsilon_T}{d p_{TS}} \varrho_{n_{TS}p_{TS}} \Delta n_{TS} \Delta p_{TS} \\ \nonumber
    & & +2 \frac{d\epsilon_{T}}{d n} \frac{d\epsilon_T}{d n_{T}} \varrho_{n_{}n_{T}} \Delta n_{} \Delta n_{T} 
    +2 \frac{d\epsilon_{T}}{d n} \frac{d\epsilon_T}{d n_{S}} \varrho_{n_{}n_{S}} \Delta n_{} \Delta n_{S} \\ \nonumber 
    & & +2 \frac{d\epsilon_{T}}{d n} \frac{d\epsilon_T}{d n_{TS}} \varrho_{n_{}n_{TS}} \Delta n_{} \Delta n_{TS} 
    +2 \frac{d\epsilon_{T}}{d n_{T}} \frac{d\epsilon_T}{d n_{S}} \varrho_{n_{T}n_{S}} \Delta n_{T} \Delta n_{S} \\ \nonumber 
    & & +2 \frac{d\epsilon_{T}}{d n_{T}} \frac{d\epsilon_T}{d n_{TS}} \varrho_{n_{T}n_{TS}} \Delta n_{T} \Delta n_{TS} 
    +2 \frac{d\epsilon_{T}}{d n_{S}} \frac{d\epsilon_T}{d n_{TS}} \varrho_{n_{S}n_{TS}} \Delta n_{S} \Delta n_{TS} \\ \nonumber 
    & & +2 \frac{d\epsilon_{T}}{d p} \frac{d\epsilon_T}{d p_{T}} \varrho_{p_{}p_{T}} \Delta p_{} \Delta p_{T} 
    +2 \frac{d\epsilon_{T}}{d p} \frac{d\epsilon_T}{d p_{S}} \varrho_{p_{}p_{S}} \Delta p_{} \Delta p_{S} \\ \nonumber 
    & & +2 \frac{d\epsilon_{T}}{d p} \frac{d\epsilon_T}{d p_{TS}} \varrho_{p_{}p_{TS}} \Delta p_{} \Delta p_{TS} 
    +2 \frac{d\epsilon_{T}}{d p_{T}} \frac{d\epsilon_T}{d p_{S}} \varrho_{p_{T}p_{S}} \Delta p_{T} \Delta p_{S} \\ \nonumber 
    & & +2 \frac{d\epsilon_{T}}{d p_{T}} \frac{d\epsilon_T}{d p_{TS}} \varrho_{p_{T}p_{TS}} \Delta p_{T} \Delta p_{TS} 
    +2 \frac{d\epsilon_{T}}{d p_{S}} \frac{d\epsilon_T}{d p_{TS}} \varrho_{p_{S}p_{TS}} \Delta p_{S} \Delta p_{TS} \\ \nonumber 
\end{eqnarray}
and equivalently for the fake rate $f_T$ and the other solved quantities.
The correlation coefficient $\varrho$ of two variables $n_a$ and $p_b$ can be determined 
by means of the relation
\begin{equation}
  \varrho_{n_ap_b} = \frac{1}{n\cup p}(n_a\cap p_b -\frac{n_ap_b}{n\cup p})
\end{equation}
where $n\cup p$ are the jets belonging to the sample $n$ or $p$ and $n_a\cap p_b$ are the jets
in common between the subsamples $n_a$ and $n_b$. 
It should be noted that the total number of jets per event could also be exploited in the definition of the 
correlation coefficient.
Given the fact that the correlation terms are in general small the definition above is used in the following
without further discussion.
To keep the numerical evaluation efficient, the derivatives are computed exploiting differentials 
($d\epsilon_T$, $df_T$, $d\epsilon_S$, $df_S$, $dn_b$, $dn_q$, $dp_b$ and $dp_q$).
 The derivative of the tagging efficiency with respect to the data sample size 
$n$ is then given by
\begin{eqnarray} 
\frac{d \epsilon_T}{dn} & = & \frac{\partial\epsilon_T}{\partial a_{00}}\frac{d a_{00}}{d n} + \frac{\partial\epsilon_T}{\partial a_{01}}\frac{d a_{01}}{d n} + \frac{\partial\epsilon_T}{\partial a_{10}}\frac{d a_{10}}{d n} + \frac{\partial\epsilon_T}{\partial a_{11}}\frac{d a_{11}}{d n} \\ \nonumber
& & + \frac{\partial\epsilon_T}{\partial b_{00}}\frac{d b_{00}}{d n} + \frac{\partial\epsilon_T}{\partial b_{01}}\frac{d b_{01}}{d n} + \frac{\partial\epsilon_T}{\partial b_{10}}\frac{d b_{10}}{d n} + \frac{\partial\epsilon_T}{\partial b_{11}}\frac{d b_{11}}{d n}    
\end{eqnarray}
and similarly for the other variables. Some of the right hand terms vanish, 
but most of them are always different from zero. Which terms vanish depends on the solved variable and 
the derivative which is being considered.
To determine the error on a subsequent quantity like the tagging efficiency of the 
auxiliary tagger $S$ (eq. \ref{epsilonSsol}) accurately, its dependences on previously 
solved-for quantities are substituted to obtain a function of the form
\begin{equation}
 \epsilon_S = \epsilon_S(a_{00}, a_{01}, a_{10}, a_{11}, b_{00}, b_{01}, b_{10}, b_{11}) .
\end{equation}
The derivatives can then be determined as above to compute the final error in this quantity.
This way of determining the errors has also the advantage that their magnitude does not depend on 
the order in which the unknowns have been solved.

\section{Performance}
\label{performance}

The system of equations does not address charm flavour sample content. Other methods
like fitting of muon in jets relative transverse momentum sample distributions to 
templates of different jet flavour contents ($b$, $c$ and light) rely on simulation 
and the $c$ and light flavour content templates are typically difficult to distinguish
\cite{Greder2004}. Within the system of eight equations the $c$ flavour content 
is predominantly absorbed into the light flavour subsample.
% with the dominating fraction 
%identified as light like by the taggers.
%A long-lived subsample which is attributed to the heavy flavour part of the sample 
%composition and a short lived one which is attributed to the light flavour part.
In this way the method is able to disentangle self-consistently the heavy and light flavour 
content of two given samples. 
However it is not possible to distinguish between light and $c$ flavour contents.
More details about this problem can be found in 
\cite{Scanlon2006}.

As a concrete example lets consider two data samples $n$, $p$ and their corresponding 
tagged subsamples
\begin{eqnarray} \label{initcond} \nonumber
  n_{\phantom{TS}} & = & 758925 , \\ \nonumber
  n_{T\phantom{S}} & = & 73076 , \\ \nonumber
  n_{S\phantom{T}} & = & 376891 , \\ 
  n_{TS} & = & 49810 , \\ \nonumber
  p_{\phantom{TS}} & = & 11082 , \\ \nonumber 
  p_{T\phantom{S}} & = & 2406 , \\ \nonumber
  p_{S\phantom{T}} & = & 7198 , \\ \nonumber
  p_{TS} & = & 1778 .
\end{eqnarray}

The solution of the unknown quantities and their errors, determined as explained in the last section,
are given by
\begin{eqnarray} \nonumber
\epsilon_T & = & 0.298 \pm 0.032 , \\ \nonumber
f_T & = & 0.026 \pm 0.005 ,\\ \nonumber
\epsilon_S & = & 0.751 \pm 0.007 , \\ \nonumber
f_S & = & 0.408 \pm 0.016 , \\
n_b & = & 195691 \pm 25645 , \\ \nonumber
n_q & = & 563234 \pm 24670 , \\ \nonumber
p_b & = & 7794 \pm  978 , \\ \nonumber
p_q & = & 3288 \pm 986 .
\end{eqnarray}
This example shows typical numbers of a life time based tagger $T$ to be probed
and an auxiliary de-correlated tagger $S$, needed to solve the system of equations.
The denominators 
\begin{eqnarray} \nonumber
 n_T - n f_T & = & 53344 \\ 
 n_T - n \epsilon_T & = & 73076 \\ \nonumber
 \epsilon_T - f_T & = & 0.272
\end{eqnarray}
are far enough from singularities to ensure
reasonably small errors on the solved quantities.
\begin{figure}[b]
\hspace*{4ex}
\includegraphics[width=16cm]{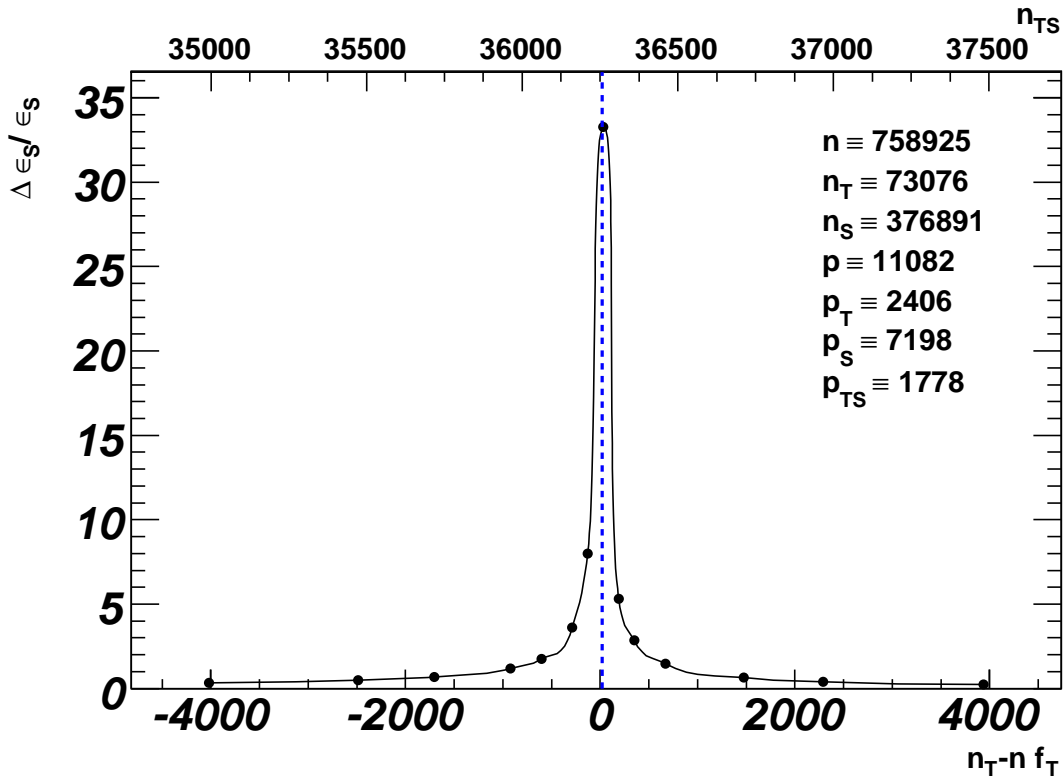}
\vspace*{-5ex}
\caption{ \label{singularity}
The relative error of the tagging rate $\epsilon_S$ as a function of the
denominator $n_T-nf_T$, which is needed for the computation of
$\epsilon_S$. The scan over the singularity has been achieved in 
varying the known quantity $n_{TS}$. Each dot corresponds to a solution
of $n_{TS}$. The interpolation function has been obtained with cubic splines.
}
\vspace*{2ex}
\end{figure}
If initial conditions have been chosen such that the solutions are getting close to 
a singularity, the computed errors diverge,
indicating that an ill-posed problem is attempted to be solved.
In this way one is protected against the misuse of poorly adjusted working points
of the taggers and badly chosen data samples.
%This will be demonstrated in an explicit example below.
To minimize the errors it is convenient to use working points of the tagging 
algorithms which are far away from the singularities.
In addition the heavy flavour content of the two data samples should not be too similar to 
each other, which would be the case if their heavy flavour contents would
be consistent with each other within errors.
In practice the two data samples can be realized by means of 
two different trigger requirements, one for each sample. An appropriate choice 
are jet only triggers on the one hand and muon plus jets triggers on the other hand. 
Since muons in jets indicate the presence of heavy flavour decays, the later category
of triggers will provide a sample enhanced in heavy flavour content with respect to the
former one. 

To demonstrate the behaviour of initial conditions approaching a singularity
the input values of the example given by eqs. (\ref{initcond}) have been taken, with exception
of the number of double tagged jets $n_{TS}$ which has been varied.
Fig. \ref{singularity} shows the relative error $\Delta\epsilon_S/\epsilon_S$
as a function of the denominator ($n_T-nf_T$) on which the tagging efficiency
$\epsilon_S$ depends. The values of the varied variable $n_{TS}$ are also indicated.
The relative error $\Delta\epsilon_S/\epsilon_S$ 
exceeds unity while approaching the singularity.
The relative error is monotonically decreasing
from the singularity to the edges of the allowed phase space $0 < n_{TS} < \min (n_T,n_S)$.
Pseudoexperiments have been conducted to verify that the computed efficiency $\epsilon_S$
is also gaussian distributed in the vicinity of the singularity around two different central
values 
%$n_{TS}\simeq35945$ 
$n_{TS}=35945$ 
for which $\Delta\epsilon_S/\epsilon_S\simeq 1$ and 
%$n_{TS}\simeq36200$ 
$n_{TS}=36200$ 
for which $\Delta\epsilon_S/\epsilon_S\simeq 10$. 
%In this way it is guaranteed that the gaussian error propagation taking correlations 
%into account is still valid in this region of phase space.
Furthermore the determined errors of the tagging efficiency and fake rate 
of the probed tagger have been checked for coverage by means of 
pseudoexperiments where the known quantities 
have been varied within their errors (see e.g. eqns. \ref{knownerrors} for the errors
of the sample $n$ and corresponding subsamples).

Different working points of the taggers can be probed to maximize the 
denominators and minimize the errors
on the most important quantities of interest,
which are the efficiency and fake rate of the probed tagger.
Their errors enter among others into the systematics of cross section 
measurements and limit calculations of searches for new physics where
heavy flavour jets are involved.

\section{Conclusions}
An algebraic way of determining heavy and light flavour jet tagging efficiencies
has been discussed. 
The analytical solution of a system of eight non-linear
equations has been obtained by means of resultants. Its singularities suggest
prescriptions to prevent badly chosen sample flavour compositions and working points
of the used tagging algorithms. Errors are obtained by gaussian error 
propagation, taking correlations between the samples and subsamples into account. 
They diverge as one approaches the singularities
making the method robust 
against its usage at badly chosen working points 
and allowing for optimisation.

\section*{Acknowledgements}
Thanks to many colleagues of the D\O\ collaboration for useful 
discussions.
This work has been supported by BMBF, DFG, a Marie Curie Early Stage Research Training Fellowship
of the European Community's Sixth Framework Programme under contract number MRTN-CT-2006-035606
and by the {\it Commissariat \`a l'Energie Atomique} and
CNRS/{\it Institut National de Physique Nucl\'eaire et de Physique des
Particules}, France.

            % Literatur

\end{document}